\documentclass[epjCONF]{svjour}
\usepackage{graphics}
\usepackage[varg]{txfonts} % Times fonts
\usepackage[latin1]{inputenc}
\session-title{Hadron Collider Physics Symposium 2011}
\begin{document}
\title{Soft Physics at RHIC}
\author{Michal \v{S}umbera\thanks{\email{sumbera@ujf.cas.cz}} (for the STAR and PHENIX Collaborations)} 
\institute{Nuclear Physics Institute ASCR, 25068 \v{R}e\v{z}, Czech Republic}
\abstract{Recent soft physics results from collisions of ultra-relativistic nuclei at Relativistic Heavy Ion Collider (RHIC) operating  at Brookhaven National Laboratory (BNL) are reviewed. Topics discussed cover the Beam Energy Scan program with some emphasis on anisotropic particle flow.} 
%end of abstract
%
\maketitle
\section{Introduction}
\label{intro}

At sufficiently high temperature $T$ or baryon chemical potential $\mu_{B}$ QCD predicts a phase transition from hadrons to the plasma of its fundamental constituents -- quarks and gluons. Search for and understanding of the nature of this transition has been a long-standing challenge to high-energy nuclear and particle physics community. In 2005, just five years after start up of RHIC, first convincing arguments on the existence of de-confined partonic matter were published \cite{Whitepaper}.  Exciting discoveries made by four experiments BRAHMS, PHOBOS, PHENIX and STAR on perfect quark-gluon liquid \cite{Whitepaper}, constituent number scaling of particle flow \cite{ncq-scaling,phi_flow}, jet quenching \cite{jetsup} and heavy-quark suppression \cite{hq} were recently complemented by the first detection of anti-strange nucleus \cite{antihyper} and by the observation of the heaviest anti-nucleus -- $^4\overline{He}$ \cite{antialpha}.  The medium produced in collisions of ultra-relativistic nuclei at RHIC,  having a highly non-trivial properties of strongly interacting quark-gluon plasma (sQGP), is definitely worth to study  over much  broader energy range.  A central goal now is to map out as much as possible of the QCD phase diagram in $T$, $\mu_{B}$  plane trying to understand various ways in which the hadron-to-QGP transition may occur. 

While the soft physics results from the high energy frontier, the Large Hadron Collider (LHC) at CERN, are covered by P.~Kuijer's contribution to  this workshop \cite{Kuijer}, the low energy frontier of RHIC is presented in this talk.  For the topics not included or not sufficiently covered in depth in this minireview I refer interested reader to consult PHENIX and STAR contributions in recently published proceedings of the  {\it Quark Matter 2011} conference \cite{QM2011}.

\section{Beam Energy Scan Program}
\label{sec:1}

During eleven years of its operation the RHIC machine has delivered a variety of nuclear beams (Au, Cu, d). The most frequently used Au+Au collisions were until recently studied  at c.m.s. energies per nucleon-nucleon pair $\sqrt{s_{NN}}$ = 200  and 62 GeV.   The last few years have witnessed, quite naturally, a shift of experimental activity from the top to lower energies. After few small-statistic exploration Au+Au runs at $\sqrt{s_{NN}}$ =22 GeV  in 2005 and  at $\sqrt{s_{NN}}$ = 9.2 GeV in 2008, two remaining  running experiments, PHENIX and STAR, collected large-statistics  data sets at  $\sqrt{s_{NN}}$ = 7.7, 11.5 and 39 GeV in 2010, and 19.6 and 27 GeV in 2011.  It is noteworthy that $\sqrt{s_{NN}}$ = 7.7GeV, which is much below the RHIC design injection energy of 19.6 GeV, is also the lowest ever achieved energy of hadron collider. The goal of this Beam Energy Scan (BES) program \cite{BES} is to search for the QCD critical point, onset of signature of QGP, and softening of the equation of state.  The QCD critical point is a distinct singular feature of the phase diagram in a $T$, $\mu_{B}$ plane, where the nature of the transition changes from a discontinuous (first-order) transition to an analytic crossover. Latter, according to lattice calculations, occurs when $\mu_{B}\approx 0$ and drives the de-confining phase transition at the top RHIC energy and above.  

\vspace*{-.6cm}
\begin{figure}[hbt]
 \begin{center}
\resizebox{1.\columnwidth}{!}{
\includegraphics{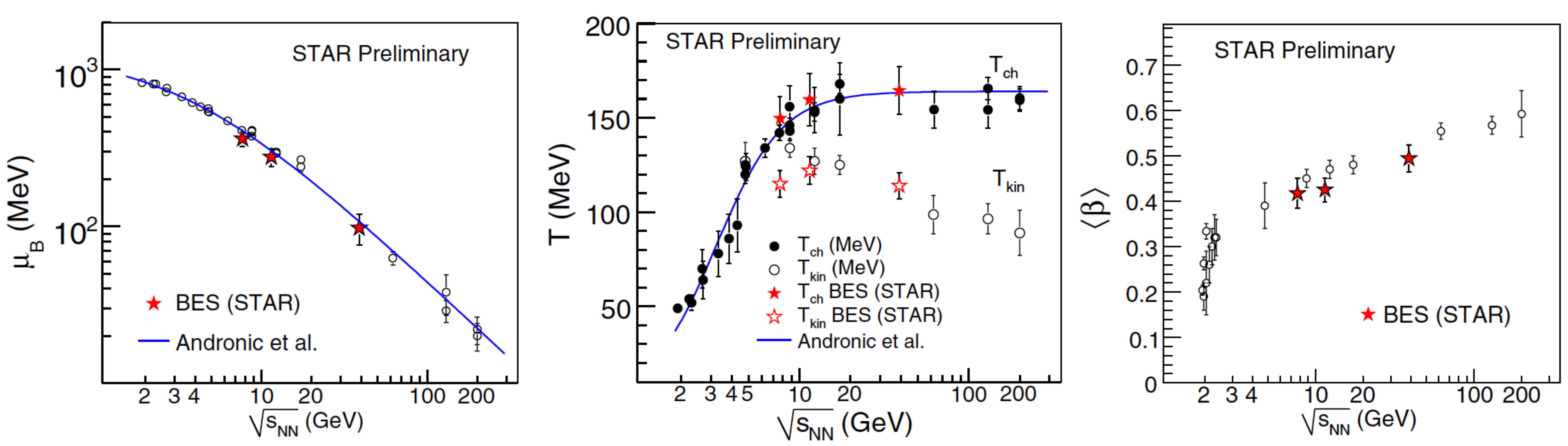}}
\end{center}
\vspace*{-.4cm}
\caption{Energy-dependence of $\mu_{B}$ (left), $T_{ch}$ and $T_{kin}$ (middle) and  $\left <\beta\right>$. The red stars are new STAR results at midrapidity from 5$\%$ most central Au+Au collisions at BES energies  and the solid lines are the model calculations \cite{Kumar}. The black symbols are previous results from the RHIC, SPS and AGS experiments \cite{BES}.}
\label{fig:1}       % Give a unique label
\end{figure}

\vspace*{-.3cm}
Statistical hadronization model fit  to mid-rapidity particle ratios ($\pi^{-}/\pi^{+}, K^{-}/K^{+},\bar{p}/p, K^{-}/\pi^{-}$ and $\bar{p}/\pi^{-}$) from 5$\%$ most central Au+Au collisions  was used by STAR to extract the chemical freeze-out (vanishing inelastic collisions) conditions \cite{Kumar}.  Fig.\ref{fig:1} shows that the BES program has extended the $\mu_{B}$ range at the RHIC from around 20MeV to about 400 MeV. $\pi^{-}, K^{-}$ and $\bar{p}$ yields were fit with a blast wave model to extract the kinetic freeze-out (vanishing elastic collisions) conditions \cite{Kumar}. The kinetic freeze-out temperature ($T_{kin}$) is observed to slightly decrease whereas the collective radial flow velocity $\left <\beta\right>$ increases with decreasing $\sqrt{s_{NN}}$. The large $\mu_{B}$ values at midrapidity indicate the formation of high net-baryon density matter, which is expected to reach a maximum value around 8 GeV \cite{BES}.

\vspace*{-.4cm}
\subsection{Anisotropic flow}
\label{sec:1}
Study of the conversion of coordinate space anisotropies into momentum space anisotropies plays a central role in ongoing efforts to characterize  the transport properties of sQGP.  The  azimuthal anisotropic flow strength is usually parametrized via Fourier coefficients $v_{n} \equiv \left< \cos\left[ n (\phi - \Psi_{n})  \right]\right>$, where $\phi$ is the azimuthal angle of the particle, $\Psi_{n}$ is the azimuthal angle of the initial-state spatial plane of symmetry (the reaction plane) and $n$ is the order of the harmonic.  The event planes from the higher moments at various rapidities are defined with various reaction plane detectors (e.g. Reaction Plane  and Muon Piston Calorimeter  detectors at  $|\eta|$ =1.0-2.8 and $|\eta|$=3.1-3.7, respectively, in PHENIX).  

%\vspace*{-.6cm}
\begin{figure}[hbt]
 \begin{center}
\resizebox{.8\columnwidth}{!}{
\includegraphics{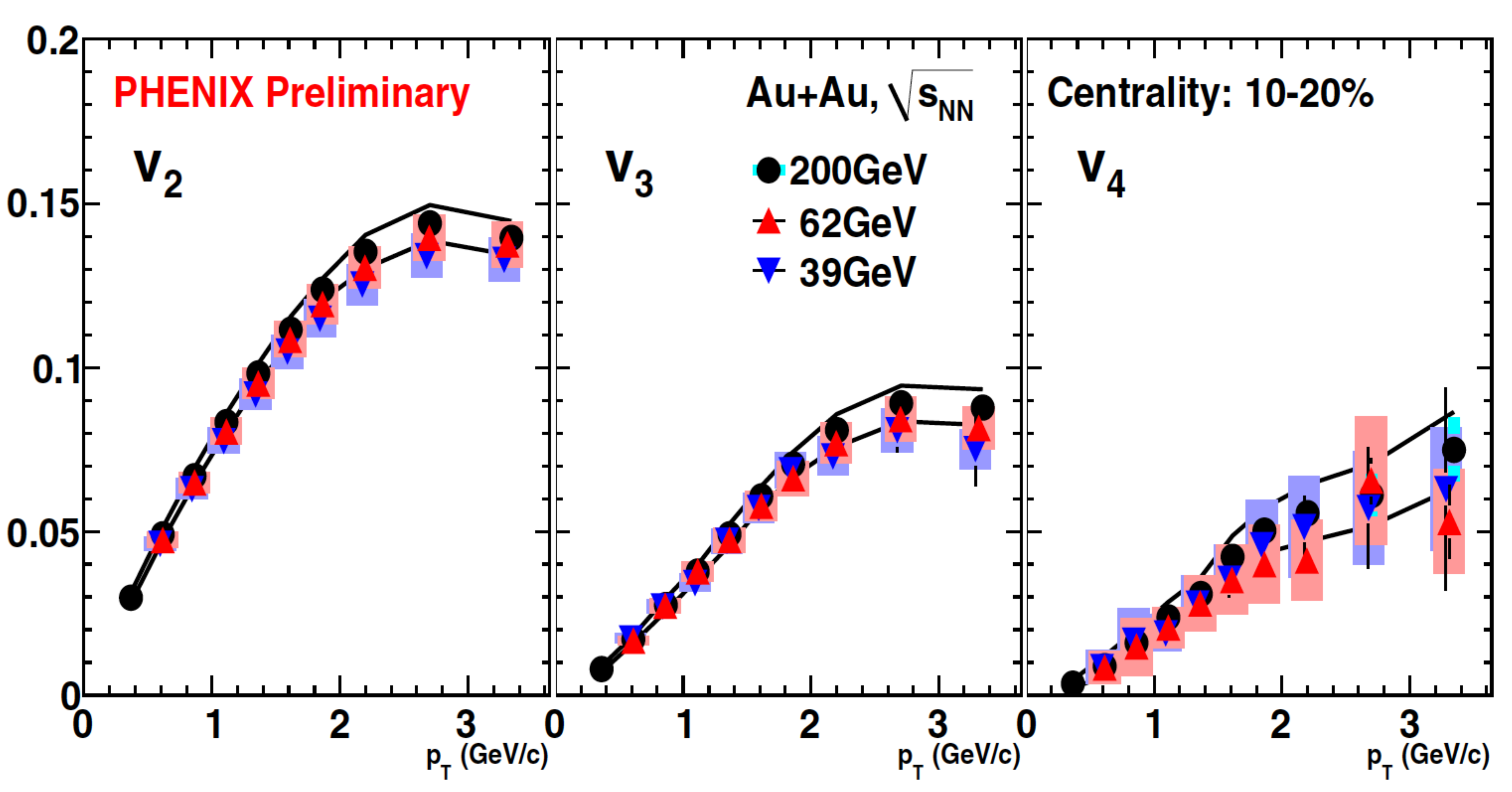}}
\end{center}
\vspace*{-.4cm}
\caption{Azimuthal asymmetry coefficients $v_{2,3,4}(p_{T})$ of charged hadrons  at mid-rapidity ($|\eta|<0.35$) from centrality 10--20$\%$ Au+Au collisions at $\sqrt{s_{NN}}$=39, 62 and 200 GeV \cite{PHENIX_V2_EDEP}. Black lines indicate systematic errors on $v_{2}$. The systematic errors for $v_{3}$ and $v_{4}$  are indicated by bands on data points. }
\label{fig:2}       % Give a unique label
\end{figure}

%\vspace*{-.5cm}
The big surprise at RHIC came from the measurement of the $v_{2}$ coefficient, integrated elliptic flow,  which brings information on the pressure and stiffness of the equation of state during the earliest collision stages.  It was found that $v_{2}$ increases by 70\% from the  top SPS energy $\sqrt{s_{NN}}$ =17.2 GeV to the top RHIC energy $\sqrt{s_{NN}}$ =200 GeV \cite{Whitepaper}.  The large value of $v_{2}$  observed at RHIC and recently also at LHC \cite{Kuijer},\cite{LHC_Flow} is one of the cornerstones of the perfect liquid bulk matter dynamics.  Moreover,  the differential $v_{2}(p_{T})$, characterizing in detail the hydrodynamic response to the initial geometry,  seems to be unchanged between the top RHIC energy and LHC energy of $\sqrt{s_{NN}}$=2.76 TeV \cite{Kuijer,LHC_Flow}. Hence, both at RHIC and at the LHC created matter behaves as the strongly coupled nearly perfect fluid. The latest results on $v_{2}(p_{T})$ shown on the left panel of Fig.\ref{fig:2}  allow us to conclude that the interval over which the elliptic flow saturates now extends almost two orders of magnitude: from 2.76 TeV  to 39 GeV. Since $v_{2}(p_{T})$ at the top SPS energy is much below the saturation curve it would be interesting to see what happens at already collected  $\sqrt{s_{NN}}$=27  and 19.6 GeV BES energies. 

At midrapidity smooth distribution of the matter in the overlapping region of two equal-mass incoming nuclei  implies vanishing of all odd harmonic.  The central panel of Fig.\ref{fig:1} shows  that, due to fluctuations in the initial matter distribution, this assumption is ill-founded. Moreover, for 39 GeV $\le \sqrt{s_{NN}}\le$ 200 GeV the data on $v_{3}(p_{T})$  seems to saturate and so the 'lumpiness' of initial geometry over this energy interval remains the same.  Excitation function of $v_{4}(p_{T})$, which could provide additional constraints on initial geometries and transport coefficients, plotted on the right panel of Fig.\ref{fig:1},  shows the similar saturation. It is noteworthy that the initial state fluctuations also show up in two-particle correlation function ${\rm\Delta}\eta$ and ${\rm\Delta}\phi$ for particles with 2 $<$ $p_{T}$ $<$ 5 GeV/$c$ from 1$\%$  most central Au+Au collisions \cite{STAR_HARMONICS}.

\vspace*{-.8cm}
\subsection{Elliptic flow of identified particles}
Interestingly, the flow patterns are also reflected in the constituent quark number ($ncq$) scaling of particle identified data. Plotting $v_{2}/ncq$ versus $(m_{T}-m_{0})/ncq$ for various particle species, where $ncq$ is the number of constituent quarks of a hadron  with mass $m_{0}$ and $m_{T}-m_{0}$ is its transverse kinetic energy, one finds the data to collapse onto a single universal curve.  Suggested by parton coalescence and recombination models \cite{phi_flow}, the universal scaling  of light flavor mesons and baryons \cite {ncq-scaling}, including multi-strange baryons and $\phi$-meson \cite{phi_flow} first observed at the top RHIC energy is  now considered as an evidence of partonic collectivity of nuclear matter.   For hadrons containing the heavy quarks the situation is less clear. Contrary to substantial elliptic flow of mesons containing the heavy quarks found recently by PHENIX \cite{PHENIX_D_FLOW} the latest STAR measurements of $J/\psi$ \cite{STAR_QM} are consistent with $v_{2}\approx 0$  disfavoring thus the coalescence scenario of $J/\psi$ production from thermalized charm quarks. 

\vspace*{-.6cm}
\begin{figure}[hbt]
 \begin{center}
\resizebox{.95\columnwidth}{!}{
\includegraphics{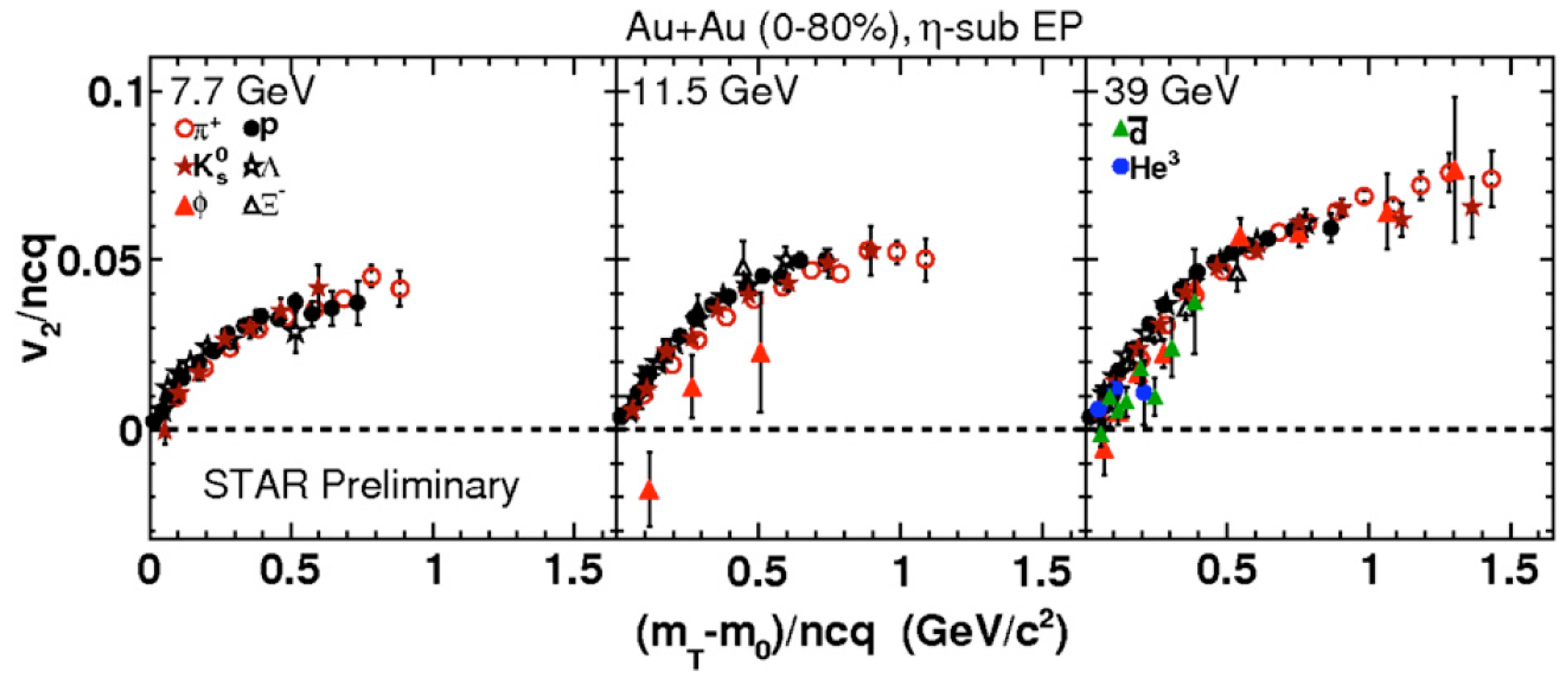}}
\end{center}
\vspace*{-.4cm}
\caption{$v_{2}/ncq$ as a function of $(m_{T}-m_{0})/ncq$ for various particles and light nuclei produced in  0-80$\%$ central Au+Au collisions  at $\sqrt{s_{NN}}$=7.7 GeV (left), $\sqrt{s_{NN}}$=11.5 GeV (middle) and $\sqrt{s_{NN}}$=39 GeV (right) \cite{STAR_BES_V2}.}
\label{fig:3}       % Give a unique label
\end{figure}

\vspace*{-.3cm}
Recent PHENIX measurements \cite{PHENIX_V2_EDEP} of elliptic flow of $\pi^{\pm}$, $K^{\pm}$, $p$ and $\bar{p}$ from Au+Au collisions confirm that at $\sqrt{s_{NN}}$ = 62 and 39 GeV the $ncq$-scaling still holds, with some deviations in the $(m_{T}-m_{0})/ncq$ range of $0.2 -0.5$ GeV/$c^{2}$ especially for (anti)protons and more prominent for 39 GeV.  This observation is confirmed and further extended by the new STAR measurements of the elliptic flow of particles at $\sqrt{s_{NN}}$= 39, 11.5 and  7.7 GeV \cite{BES}.  Differences are observed in the $v_{2}$ of particles and anti-particles, which increase as $\sqrt{s_{NN}}$ decreases suggesting that the $ncq$-scaling for all particle species (including nuclei) as observed at top RHIC energies \cite{phi_flow} is no longer valid at these lower energies.

Fig.\ref{fig:3} shows the STAR data on elliptic flow of various particles produced in 0-80$\%$ central Au+Au collisions at $\sqrt{s_{NN}}$=7.7, 11.5 and 39 GeV \cite{STAR_BES_V2}.  Most of the particle species follow the $ncq$-scaling, except for the $\phi$-mesons, which have $v_{2}$ at 11.5 GeV systematically lower  than the other hadrons. This may provide an evidence for a change in the degrees of freedom around $\sqrt{s_{NN}} \approx$ 10 GeV.  If in addition, a hierarchy of the violation of the $ncq$-scaling could be established when going from $p$ to $\Lambda$, $\Xi$ and $\Omega$ it could provide further insights into the relative importance of hadronic and partonic phase  in the early stage of the reaction.

\section{Summary}
Recent soft physics results from RHIC have substantially extended our knowledge of hot and dense de-confined QCD matter.  The BES program covering a large part of the conjectured QCD phase diagram revealed significant differences in particle and anti-particle $v_{2}$ coming from the high net-baryon density at midrapidity. Small $v_{2}$ of $\phi$-meson indicates that hadronic interactions start to dominate over partonic interactions around 11.5 GeV.  Saturation of differential elliptic flow $v_{2}(p_{T})$ from $\sqrt{s_{NN}}$ = 2.76 TeV down to $\sqrt{s_{NN}}$ = 39 GeV extends substantially the region where the sQGP can be created and studied under controlled laboratory conditions. A non-negligible contribution to azimuthal anisotropy of produced particles comes from the fluctuations in the initial matter distribution of colliding nuclei.

\section*{Acknowledgements}
This work was supported in part by grants LC07048 and LA09013 of the Ministry of Education of the Czech Republic.

\end{document}